\documentclass[prd,aps,showpacs,nofootinbib]{revtex4}
\usepackage{graphicx}
\usepackage{amsfonts}
\usepackage{amssymb}
\usepackage{latexsym}
\newcommand{\CL}{{\cal L}}
\newcommand{\CM}{{\cal M}}

\newcommand{\CD}{{\cal D}}

\newcommand{\bear}{\begin{array}}  \newcommand{\eear}{\end{array}}
\newcommand{\bea}{\begin{eqnarray}}  \newcommand{\eea}{\end{eqnarray}}
\newcommand{\beq}{\begin{equation}}  \newcommand{\eeq}{\end{equation}}
\newcommand{\bef}{\begin{figure}}  \newcommand{\eef}{\end{figure}}
\newcommand{\bec}{\begin{center}}  \newcommand{\eec}{\end{center}}
\newcommand{\non}{\nonumber}  
\newcommand{\lmk}{\left(}  \newcommand{\rmk}{\right)}
\newcommand{\lkk}{\left[}  \newcommand{\rkk}{\right]}
\newcommand{\lhk}{\left \{ }  \newcommand{\rhk}{\right \} }
\newcommand{\lnk}{\left \{ }  \newcommand{\rnk}{\right \} }
  
\newcommand{\del}{\partial}  
\newcommand{\vect}[1]{\mbox{\boldmath${#1}$}}
\newcommand{\vecs}[1]{\mbox{\boldmath\tiny${#1}$}}
\newcommand{\bib}{\bibitem} \newcommand{\new}{\newblock}
\newcommand{\la}{\left\langle} \newcommand{\ra}{\right\rangle}

\newcommand{\GeV}{{\rm GeV}}

\newcommand{\Tr}{\mbox{Tr}}
\newcommand{\phip}{\phi_{+}}
\newcommand{\phim}{\phi_{-}}

\newcommand{\phid}{\phi_{\Delta}}

\newcommand{\im}{\mbox{Im}}
\newcommand{\re}{\mbox{Re}}

\newcommand{\phik}{\phi_{\vecs k}}
\newcommand{\Cak}{C_{\vecs k}}

\newcommand{\xik}{\xi_{\vecs k}}
\newcommand{\vek}{\vect k}
\newcommand{\vep}{\vect p}
\newcommand{\vex}{\vect x}

\newcommand{\omp}{\omega_{p}}

\def\PRD#1#2#3{Phys. Rev. D {\bf #1}, #2 (19#3)}

\begin{document}
\title{Can oscillating scalar fields decay into particles with a 
large thermal mass?}

\author{Jun'ichi Yokoyama} 
\affiliation{Research Center for the Early Universe (RESCEU), Graduate
School of
Science, The University of Tokyo, Tokyo 113-0033, Japan}
%%%

\date{\today}
%%

%\maketitle

\begin{abstract}
We calculate the
 dissipation rate of a coherently oscillating scalar field 
 in a thermal environment 
using nonequilibrium quantum field theory and apply it to 
the reheating stage after cosmic inflation.
It is shown that the rate is nonvanishing  even when
particles coupled to the oscillating inflaton field
 have a larger thermal mass than it, 
and therefore the cosmic temperature can be much higher
than inflaton's mass even in the absence of preheating.
Its cosmological implications are also discussed.

\end{abstract}

\pacs{98.80.Cq,11.10.Wx,05.40.-a \hspace{1.5cm} RESCEU-15/05}

\maketitle

%\newpage
\tighten

\section{Introduction}

In our contemporary understanding, the origin of the primeval fireball
whose existence was assumed in the conventional hot big bang cosmology
is the reheating processes after inflation---an accelerated cosmic
expansion which has made the universe
  homogeneous and spatially flat
 with small density fluctuations that eventually grow to the
observed large-scale structure \cite{Sato:1980yn,lindebook}.
The universe is reheated through the dissipation
 of coherent oscillation of
the zero-mode of the {\it inflaton}, the scalar field whose potential
energy drives inflation.  While the initial stage of reheating could 
be rather complicated due to an explosive particle production induced 
by parametric resonance, which is dubbed as preheating
\cite{Traschen:1990sw}, the final
stage is dominated by perturbative decay. The latter process
 determines the reheat
temperature, $T_R$, the temperature at the outset of the radiation 
domination \cite{Albrecht:1982mp}.

Note, however, that in general $T_R$ is much lower than 
the highest temperature the universe has ever
experienced after inflation even in the case only perturbative decay
operates to reheat the universe \cite{Giudice:2000ex}.
This means that in the late stage of the reheating processes, the 
inflaton decays not in a vacuum but in a thermal medium.  
About this point an interesting
claim has been made in \cite{Kolb:2003ke}  that if the 
would-be decay products of the oscillating
inflaton  acquire a thermal mass larger than the inflaton
mass in the thermal background, it cannot decay into these
particles, and that reheating is suspended for some time, based on the
observation that the phase space would be closed for the mass of the
decay product being larger than half the inflaton mass.  
The decay width of the inflaton $\phi$ with mass $m_\phi$
into two massive particles with
mass $m$ reads
\beq
  \Gamma_\phi=\Gamma_{\phi 0}\lmk 1-\frac{4m^2}{m_\phi^2}\rmk^{1/2},
\eeq
where $\Gamma_{\phi 0} $ is the decay rate in the case $m=0$.
So if we simply replace $m$ with a thermal mass $m(T)\sim gT$ and if it
is larger than $m_\phi/2$, the phase space is closed and inflaton decay
is apparently forbidden. Here $g$ is some coupling constant of the
would-be decay product.  Then thermal history after inflation would 
be drastically changed.  That is,
the highest temperature in this era
cannot exceed $\sim m_\phi/g$ if preheating is inoperative, and also the 
reheat temperature is bounded from above 
by $m_\phi/g$ and is independent of
the decay rate of the inflaton in case the conventional calculation 
gives a larger value.
The former would change the abundance of supermassive particles and 
the latter affects the gravitino abundance \cite{Kolb:2003ke}, because
the gravitino-entropy ratio after inflation is proportional to the
temperature at the onset of radiation domination.

Furthermore,
this situation is not specific to the reheating stage after inflation
but may apply in any epoch when significant amount of entropy is 
produced out of the decay of oscillating scalar field with a relatively 
small mass.
Indeed the above 
possibility was first pointed out by Linde \cite{Linde:gh}
in the context of Affleck-Dine baryogenesis \cite{Affleck:1984fy}
where the Affleck-Dine scalar field oscillates with a mass of order
of $10^{2-3}$ GeV in a medium with a much higher temperature.  The final
magnitude of baryon asymmetry changes if this suspension of decay
is operative \cite{Linde:gh}.

The above naive picture, however, may be too simplistic
because a thermal mass is different from the intrinsic mass and because
coherent field oscillation is different from a collection of particles.
Thus it is very important to analyze this problem from a more
fundamental point of view, for it has a profound implication not
only to the cosmology of the early Universe but also to particle physics
in that it strongly affects various 
species of the particles produced in the early universe
 as mentioned above.
In this paper, extending our previous work \cite{Yokoyama:2004pf},
we analyze this problem in terms of 
 a nonequilibrium quantum field theory
at finite temperature \cite{Chou:es}.  Inclusion of 
 thermal masses of the would-be decay products of 
the inflaton  is achieved by adopting a resummed propagator when
we calculate the effective action for $\phi$.  
As a result of resummation
the self energy of the decay product acquires not only real part,
which appears as a high-temperature correction to the mass, but also an
imaginary part.  The latter plays a crucial role in determining the 
dissipation rate of the inflaton.  Consequently, we find that the
inflaton can dissipate its energy even when its would-be decay products
have a larger thermal mass than the inflaton itself.

\section{Model and equation of motion}

For clarity we adopt a simpler model than \cite{Yokoyama:2004pf},
that is, we adopt the following Lagrangian.
\beq
   \CL = \frac12\,(\del_{\mu}\phi)^2-\frac12\,m_{\phi}^2\phi^2 
        + \frac12\,(\del_{\mu}\chi)^2-\frac12\,m_{\chi}^2\chi^2 -
    \CM \phi\chi^2 - 
     \frac{1}{4}\,g^2 \chi^4 \:.  \label{lagrangian}
\eeq
Here $\phi$ is an oscillating real scalar field and $\chi$ is another
real scalar field.  $\phi$ can decay into a pair of $\chi$ particles 
through the interaction $\CM \phi\chi^2$, if
it is energetically allowed.  The dimensionful coupling constant
$\CM$ may be written as $\CM=hm_{\phi}$ in some supersymmetric 
inflation models
where $h$ is a Yukawa coupling 
\cite{Murayama:1992ua}.  In such models $\phi$ can also decay
into two fermions, which is suppressed by Pauli-blocking at finite
temperature.  On the contrary, the decay into two bosons is enhanced
due to the 
induced emission.  This is the reason we consider the latter
decay process. 

We analyze the behavior of the above system under the following
assumption which mimic the cosmological situations we are interested in,
such as the late reheating phase after inflation.  First we neglect
cosmic expansion since we are interested in the phenomena which
occur in a shorter time scale than the expansion time.  Second we 
assume $\chi$ is in a thermal state with a specific temperature
$T=\beta^{-1}$ and that it acquires a large thermal mass due to 
the self coupling. Note that $\chi$ can easily be thermalized during the
reheating stage since its thermalization rate, $\sim g^4T$, can
naturally  be much larger than the cosmic expansion rate.
 Finally the scalar field $\phi$ is oscillating 
but we consider the situation the parametric resonance is already
terminated with a field amplitude $\CM |\phi|< m_\phi^2 $.

Due to its coherent nature, scalar field oscillation behaves nearly
classically, but its decay is of course a quantum process.  So we 
calculate an effective action for $\phi$ and derive an equation of
motion for its expectation value.  For this purpose we 
should use the in-in or the closed time-path 
formalisms in which 
the time contour starting from the infinite past must run 
to the infinite
future without fixing the final condition and come back to the
infinite past again in calculating the generating
functional \cite{Sch}. 
This method has been applied to various cosmological problems by a
number of authors \cite{Mor,mm,boy95}.
%Calzetta:1986ey,Boyanovsky:vi,
%GR,Boyanovsky:1994me,boy95,Boyanovsky:1996xx,
%Yamaguchi:1996dp,GM,Yamaguchi:1997sy,Yokoyama:1998ju,bdh}. 
The generating functional in the present model
is  given by
\bea
  Z[J,K]  =
          \Tr \lkk T_{-}\!\lnk
            \exp\lkk i\int_{\infty}^{-\infty}\!\!\!\!\!\! dt 
\int \!\! d^3x
(J_{-}\phi_{-}+K_{-}\chi_{-})
                    \rkk \rnk 
          T_{+}\!\lnk \exp\lkk i\int_{-\infty}^{\infty}
\!\!\!\!\!\! dt \int \!\! d^3x
                (J_{+}\phi_{+}+K_{+}\chi_{+})
               \rkk \rnk \rho\,\rkk \equiv e^{iW[\,J,K\,]}.
\eea
where  $X_{+}$ denotes a field component $X$ on the plus-branch
($-\infty$ to $+\infty$) and $X_{-}$ that on the minus-branch ($+\infty$
to $-\infty$). The symbol  $T_{+}$ represents 
the ordinary time ordering, and
$T_{-}$ the anti-time ordering. $J_{\pm}$ and $K_{\pm}$ are
the external fields for $\phi$ and $\chi$,
respectively.  $\rho$ is the initial density matrix which is assigned
according to the assumption above mentioned.

In terms of the components along the plus and the minus branches,
the effective action reads
\beq
\Gamma[\phip,\phim]=W[J_+,J_-,K_{\pm}=0]-\int_{-\infty}^{\infty}
\!\!\! dt \int \!\!
d^3x \lkk J_+(x)\phip (x) -J_-(x)\phim (x)\rkk,
\eeq
with $\phip (x)=\delta W[J_+,J_-] / \delta J_+(x)$ and
 $\phim (x)=-\delta W[J_+,J_-] / \delta J_-(x)$.

\begin{figure}[htb]
\begin{center}
\includegraphics[width=4cm]{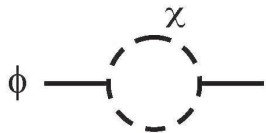}
\end{center}
\caption{One-loop 
Feynman diagram incorporated in the effective
 action.  Solid line denotes $\phi$, and broken line $\chi$.}
\label{fig:one}
\end{figure}

Here we consider a one-loop correction 
depicted in Fig.\
\ref{fig:one} which includes the essential effect in our analysis for
illustration.  We also note that instead of $\phi_+$ and $\phi_-$
it is more convenient to use $\phi_c\equiv (\phip +\phim )/2$
and $\phid\equiv \phip -\phim$ and set $\phid \longrightarrow 0$
in the end because $\phip$ and $\phim$ should be identified with
each other eventually.
Then the effective action to this order is given by
\bea
&&  \Gamma[\phi_{c},\phi_{\Delta}] =-\int d^4x %\lnk
        \phi_{\Delta}(x)( \,\Box+m_\phi^2\,)  \phi_{c}(x)
                                           \non \\
       && -\int d^4x d^4x' 
            C(x-x')
 \theta(t-t')\phi_{\Delta}(x)\phi_{c}(x') 
   +\frac{i}{2}\int d^4x d^4x' 
                D(x-x')
\phi_{\Delta}(x)\phi_{\Delta}(x')
 \:, 
  \label{eqn:a3effe}
\eea
%where
\bea
C(x-x')  \equiv 4\CM^2 \im\lkk
G_\chi^{F}(x-x')^2\rkk  \:,~~~% \label{35} \\
D(x-x')  \equiv 2\CM^2 \re\lkk G_\chi^{F}(x-x')^2\rkk \:,
\label{41}
\eea
where $G_\chi^{F}(x)$ is the Feynman propagator at finite temperature.
Its Fourier modes read \cite{pro,pro2}
\bea
 G_\chi^F(\vep,t)&=&\int \!\!d^3x\; G_\chi^F(\vect x,t)e^{-i\vecs
 p\cdot \vecs x } 
=\frac{1}{2\omp}\lnk \lkk 1+n_B(\omp)\rkk e^{-i\omp t}
+n_B(\omp)e^{i\omp t}\rnk , ~~~ \omp\equiv \sqrt{\vep^2+m_\chi^2}.
\label{bare}
\\
 G^{F}_{\chi}(p) &=& 
\int\!\! dt\; G_\chi^F(\vep,t)e^{ip_0 t}=
\frac{i}{p^2-m_{\chi}^2+i\epsilon}
                          +2\pi n_B(\omp)   
                \,\delta(p^2-m_{\chi}^2) \:,~~~
n_B(\omp)=\frac{1}{e^{\beta\omp}-1}. \label{Feynman}
\eea

The resultant effective action is complex-valued as a manifestation of
the dissipative nature of the system.  We cannot obtain any sensible
equation of motion by simply differentiating with respect to a field
variable because we are dealing with a real scalar field and its
equation of motion should be real-valued.  The cure for this problem has
been proposed by Morikawa \cite{Mor}, according which we introduce
an auxiliary random
Gaussian field, $\xi(x)$, to rewrite the effective action as follows.
\beq
  \exp (i\Gamma[\phi_{c},\phi_{\Delta}])=\int\CD\xi
              P[\xi]\exp\lhk
           i\Gamma_{\rm eff}[\,\phi_{c},\phi_{\Delta},\xi\,]\rhk
                    ,\ 
\eeq
where
%{\rm ~where~}
\beq
  \Gamma_{\rm eff}[\,\phi_{c},\phi_{\Delta},\xi\,] \equiv
                 \mbox{Re}\Gamma[\phi_{c},\phi_{\Delta}]
                  +\int d^4x\,\xi(x)\phi_{\Delta}(x) \:.
                  %  +\xi_{m}(x)\phi_{c}(x)\phi_{\Delta}(x)\,] \:.
 \label{eqn:a3cla}
\eeq
Here $P[\xi]$ is 
 a statistical distribution functional for $\xi(x)$ which is a
 Gaussian
with its dispersion given by the imaginary part of the effective action,
$\la\,\xi(x)\xi(x')\,\ra = D(x-x')$.

While the 
mathematical equivalence between the above decomposition and the
original expression (\ref{eqn:a3effe}) can easily be confirmed by
performing the path integral with respect to $\xi(x)$,  we
keep it as it is to obtain a real-valued equation of motion through 
\beq
\left.\frac{\delta\Gamma_{\rm eff}[\,\phi_{c},\phi_{\Delta},\xi\,]}
{\delta \phi_{\Delta}}\right|_{\phi_{\Delta}=0}=0.
\eeq
From (\ref{eqn:a3cla}), it reads
\bea
 (\,\Box+m_\phi^2\,)\,\phi_{c}(x)
          + \int_{-\infty}^{t}dt'\int d^3x'C(x-x')\phi_{c}(x') 
           =\xi(x) \:. \label{lineareq}
\eea
  Hereafter we omit the
suffix $c$.
The solution to the above equation of motion can be readily found
through Fourier transform,
\beq
  \phik(t)=\int d^3x \phi(\vex,t)e^{-i\vecs k\cdot\vecs x},~~~
  \tilde\phik(\omega)=\int dt \phik(t)e^{i\omega t},
\eeq
etc.  We find (\ref{lineareq}) is transformed as 
\bea
\lkk-\omega^2+\vek^2+m_\phi^2+S_{\vecs k}(\omega)\rkk\tilde\phik(\omega)
-i\omega\tilde\Gamma_k\tilde\phik(\omega)=\tilde\xik(\omega), 
\label{eqm}
\eea
where we have defined
\begin{eqnarray}
\tilde\Gamma_k(\omega)\equiv i\frac{\tilde\Cak(\omega)}{2\omega},
\label{Gammadef} ~~~
S_{\vecs k}(\omega)\equiv \int\frac{d\omega'}{2\pi}{\rm
P}\frac{1}{\omega-\omega'}i\tilde\Cak(\omega'),
\end{eqnarray}
with
\beq
 \tilde\Cak(\omega)\equiv\int dt d^3x  
C(x)e^{i\omega t-i\vecs k\cdot \vecs x },
\eeq
which is pure imaginary.  
Since $S_{\vecs k}(\omega)$ is divergent, we subtract the divergence at
$\omega=0$ to renormalize mass $m_\phi$ \cite{boy95}.  
We can show that the
renormalized part $S_{{\vecs k}\rm ren}(\omega)\equiv 
S_{\vecs k}(\omega)-S_{\vecs k}(0)$ is
of order of $\CM^2/(4\pi^2)$.  From now on we
refer to $m_\phi$ as the renormalized mass, and assume that $\CM <
m_\phi$.  Then one can neglect $S_{\vecs k}(\omega)$ in (\ref{eqm})
to yield the solution
\bea
 \phik(t) &=& \lkk\phik(t_i)\cos M_k(t-t_i)
 + \frac{\dot\phik(t_i)}{M_k}
\sin M_k(t-t_i)\rkk e^{-\frac{1}{2}\tilde\Gamma_k(M_k)(t-t_i)} \non\\
 &&+\frac{1}{M_k}\int_{t_i}^t dt' e^{-\frac{1}{2}\tilde\Gamma_k(M_k)(t-t')}
 \sin M_k(t-t')\xik(t'),~~~ M_k^2\equiv m_\phi^2+\vek^2. \label{phisol}
\eea
Here we have also 
assumed $\tilde\Gamma_k(M_k) \ll M_k$, which is justified
whenever perturbation theory applies.
From the above solution we can read off that $\tilde\Gamma_k(M_k)$
gives the dissipation rate of the field oscillation.
In particular, the dissipation rate of the zero-mode condensate
is given by
\beq
  \tilde\Gamma_0(m_\phi)=\frac{\CM^2}{8\pi m_\phi}
\lkk 1-\lmk\frac{2m_\chi}{m_\phi}\rmk^2\rkk^{1/2}
\lkk 1+2n_B\lmk\frac{m_\phi}{2}\rmk \rkk
=\Gamma_B(0)
\lkk 1+2n_B\lmk\frac{m_\phi}{2}\rmk \rkk
\simeq\frac{\CM^2T}{2\pi m_\phi^2}
\lkk 1-\lmk\frac{2m_\chi}{m_\phi}\rmk^2\rkk^{1/2} , \label{chidecayhigh}
\eeq
for $m_\phi > 2m_\chi$.  Here the last approximate
equality applies at high temperature $T\gg m_\phi$, and  
we have used
\bea
 \tilde{C}_{\vecs 0}(m_\phi)
=-i\pi\CM^2\int\!\!\!\frac{d^3p}{(2\pi)^3}
\frac{1}{\omega_p^2} (1+2n_p)\lkk\delta(m_\phi-2\omega_p)
- \delta(m_\phi+2\omega_p)\rkk,~
n_p\equiv n_B(\omp), 
\label{Cbosonzero}
\eea
for $\vek = \vect 0$ mode.  Here the delta functions
appear as a result of time-integral of the form 
\beq
\int_0^{\infty}\cos(m_\phi\mp 2\omega_p)dt=2\pi\delta(m_\phi\mp 
2\omega_p),
\eeq
which arises in turn because
the spatial Fourier mode of the
Feynman propagator at finite temperature can be expressed as
(\ref{bare}).

\section{Inclusion of the thermal mass of the decay products}

So far, although we have taken into account the fact that the decay
product $\chi$ is thermally populated, we have not considered effects
of the thermal environment on $\chi$ itself which induces a thermal mass
to $\chi$ particles.  If we simply replaced the intrinsic mass $m_\chi$
with the thermal mass $m_\chi (T)$ in (\ref{chidecayhigh}) and 
(\ref{Cbosonzero}), we would find the dissipation rate would vanish for
$m_\chi (T) > m_\phi/2$ as was claimed in \cite{Linde:gh,Kolb:2003ke}.
Here we carefully examine how to incorporate the thermal mass of $\chi$
in our analysis in order to see the validity of such a simplistic argument.
To this end, we should not rely
on the formula (\ref{chidecayhigh}) but go back to the more fundamental
equation (\ref{Cbosonzero}) to reconsider how this equation was derived.

First we note that  $\chi$'s mass in  $\omega_p$ of 
(\ref{Cbosonzero}) is that appears in the denominator of the Feynman
propagator (\ref{Feynman}) and the delta function emerges due to the
infinitesimally small imaginary part $i\epsilon$.  
The effect of the thermal environment on $\chi$, which gives rise to
finite-temperature correction to its mass, 
can be incorporated to the calculation of the dissipation rate of $\phi$
if we apply resummation and use a resummed propagator of $\chi$ 
instead of the finite-temperature bare Feynman propagator (\ref{Feynman})
to calculate the effective action.  
With the help of the Matsubara formalism \cite{Matsubara:1955ws}, 
 the denominator of the propagator acquires a self energy whose real
 part yields a finite-temperature correction to the mass as desired.
This resummation procedure, however, generates appreciable magnitude of
the imaginary part to the self energy at the same time, so that the delta
function seen in (\ref{Cbosonzero}) will no longer be present as we see below.  
Thus we expect that the use of a resummed propagator changes the result
qualitatively, and that the simple observation 
that a large thermal
mass would close the phase space of the decay rate of the inflaton 
(\ref{chidecayhigh}) would not apply.

We write real and imaginary parts of $\chi$'s self energy, $\Sigma(p)$,
 as
$\Sigma_R(p)$ and $\Sigma_I(p)$, respectively.  Then the spectral
function reads
\bea
\rho_s(\vep,\omega)&=&i\lkk 
\frac{1}{(\omega+i\epsilon)^2-\vep^2-m_\chi^2-\Sigma_R-i\Sigma_I}-
\frac{1}{(\omega-i\epsilon)^2-\vep^2-m_\chi^2-\Sigma_R+i\Sigma_I} \rkk
\non\\
&=&i\lkk 
\frac{1}{(\omega+i\Gamma_{\chi p})^2-\omega_p'^{2}}
-\frac{1}{(\omega-i\Gamma_{\chi p})^2-\omega_p'^{2}}
\rkk,
\eea
where
\bea
  \omega_p'^{2} &\equiv& \vep^2 + m_\chi^2 +\Sigma_R(p) 
 +\Gamma_{\chi p}^2(\omega) \cong \vep^2 + m_\chi^2 +\Sigma_R(p)
 ,~~  
 \Gamma_{\chi p}(\omega) \equiv -\frac{\Sigma_I(p)}{2\omega}.
\eea
The full dressed propagator is then given by
\bea
 G_{\rm full}(\vep,t)&=&\int\frac{d\omega}{2\pi}
\lnk \lkk 1+n_B(\omega)\rkk\theta(t)+n_B(\omega)\theta(-t)\rnk
\rho_s(\vep,\omega)e^{-i\omega t} \non\\
&=&\frac{1}{2\omega'_p}\lnk
\lkk 1+n_B(\omega'_p-i\Gamma_{\chi p})\rkk e^{-i\omega'_p|t|
-\Gamma_{\chi p}|t|}
+n_B(\omega'_p+i\Gamma_{\chi p})e^{i\omega'_p|t|-\Gamma_{\chi p}|t|}\rnk,
\label{full}
\eea
which should be compared with the bare propagator (\ref{bare}) 
\cite{pro2}.
Then since the propagator has a complex phase now, the simple
cosine integral in our previous calculation, which gave rise to the
delta function in (\ref{Cbosonzero}), is replaced by
\beq
  \int_0^{\infty}\!\!\! dt\; e^{-2\Gamma_{\chi p}t}
\cos(m_\phi-2\omega'_p)t=\frac{2\Gamma_{\chi p}}
{(m_\phi-2\omega'_p)^2+(2\Gamma_{\chi p})^2}, \label{BW}
\eeq
namely, the Breit-Wigner function, when we calculate (\ref{Cbosonzero})
in terms of the dressed propagator (\ref{full}).

Since (\ref{BW}) is finite even for $m_\phi \ll 2\omega'_p$
we find the dissipation rate of $\phi$ is nonvanishing even when 
$m_\phi \ll 2m_\chi (T)$.
To the first order in $\Gamma_{\chi p}$, the
 zero mode dissipation rate in such a regime is given by

\bea
  \tilde\Gamma_{\vecs 0}(m_\phi)&=&
 \frac{\CM^2}{2m_\phi}\int\frac{d^3p}{(2\pi)^3}\frac{1}{\omega_p'^2}
\lnk \frac{2\Gamma_{\chi p}(1+2n'_p)}{4\Gamma_{\chi p}^2+(m_\phi-2\omega'_p)^2}
- \frac{2\Gamma_{\chi p}(1+2n'_p)}{4\Gamma_{\chi
  p}^2+(m_\phi+2\omega'_p)^2}\right.
 \non\\
&&\left. +2\beta\Gamma_{\chi p} n'_p(n'_p+1)\lkk
\frac{2m_\phi}{4\Gamma_{\chi p}^2+m_\phi^2}-
\frac{m_\phi-2\omega'_p}{4\Gamma_{\chi p}^2+(m_\phi-2\omega'_p)^2}-
\frac{m_\phi+2\omega'_p}{4\Gamma_{\chi
  p}^2+(m_\phi+2\omega'_p)^2}\rkk\rnk, \label{resu}
\eea
where $n'_p\equiv n_B(\omega'_p)$.  
All the terms in the second line 
vanish in the limit $\Gamma_{\chi p} \longrightarrow 0$, 
while those in the
first line reduce to delta functions in the same limit and the previous
result with the undressed propagator is recovered except for the
  replacement $\omega_p\longrightarrow \omega'_p$.

As mentioned above, 
the above result remain finite even if $m_\phi$ is smaller than 
$2\omega'_p$.  In particular, if thermal mass of $\chi$ 
is much larger than the inflaton mass, $\omega'_p \gg m_\phi$,
we obtain
\bea
\tilde\Gamma_{\vecs 0}(m_\phi)&=& \frac{\CM^2}{16\pi^2}
\int_{m_{\chi}(T)}^\infty \!\!\! d\omega'_p
\frac{\sqrt{\omega_p'^2-m_{\chi}^2(T)}}{\omega_p'^4}
(1+2n'_p)\Gamma_{\chi p} +\frac{\CM^2}{\pi^2 m_\phi^2T}
\int_{m_{\chi}(T)}^\infty \!\!\! d\omega'_p
\frac{\sqrt{\omega_p'^2-m_{\chi}^2(T)}}{\omega_p'}
n'_p(n'_p+1)\Gamma_{\chi p} .  \label{oyoy}
\eea
In the present simple model, $\chi$ thermalizes only through self
interaction $g^2\chi^4/4$, then $\Gamma_{\chi p}(\omega'_p)$ is
given by
 $\Gamma_{\chi p}(\omega'_p)\cong 3g^4T^2/(128\pi \omega'_p)$ 
\cite{Hosoya:1983ke}.
We find the second term yields the dominant contribution
in (\ref{oyoy}) for $m_\phi \ll m_\chi(T) \approx gT/2$.
As a result we obtain
\beq
\tilde\Gamma_{\vecs 0}(m_\phi)\simeq \frac{\CM^2 T}{2\pi m_\phi^2}
\frac{3g^2}{24\pi^2}. \label{diss}
\eeq
Here the former factor is identical to the dissipation rate to massless
particles at high temperature $T \gg m_\phi$, and the latter factor 
represents the suppression due to the large thermal mass of the decay
product. 
One may wonder the suppression factor might be proportional to $g^4$
just as $\Gamma_{\chi p}(\omega'_p)$.  However, the result of
integration of the second term  in (\ref{oyoy}) yields $m^2_\chi(T)$
in the denominator, which partially cancels $g^4$ to $g^2$.
Although the numerical value of the suppression factor 
in (\ref{diss}) is specific to the
present model, it is a generic feature that the suppression is 
proportional to some combination of coupling constants which is
related to thermalization processes of the decay product, because
it arises from $\Gamma_{\chi p}/\omega'_p$.

\section{Discussion}

The resultant reheat temperature in the present model is given by
\beq
  T_R\approx \lmk\frac{90}{\pi^2g_*}\rmk^{1/2}
  \frac{\CM^2M_G}{2\pi m_\phi^2}\frac{3g^2}{24\pi^2}
=1.0\times 10^{11}g^2\lmk\frac{g_\ast}{200}\rmk^{-1/2}
\lmk\frac{\CM}{10^6\GeV}\rmk^2
\lmk\frac{m_\phi}{10^8\GeV}\rmk^{-2} \GeV,  \label{newreheat}
\eeq
where $g_*$ is the effective number of relativistic degree of freedom
and $M_G$ is the reduced Planck scale.
Again it is suppressed by the same factor, $3g^2/(24\pi^2)$,
compared with the case the scalar field decays into massless particles
in the high-temperature medium, see eq.\ (157) of
\cite{Yokoyama:2004pf}.  
The above result has been obtained under the assumption of 
$m_\phi \ll gT_R$ and $\CM < m_\phi$.  The former condition reads
\beq 
  m_\phi \ll 1.0\times 10^9 g\lmk\frac{g_\ast}{200}\rmk^{-1/6}
 \lmk\frac{\CM}{10^6\GeV}\rmk^{2/3} \GeV.
\eeq
 The above reheat temperature (\ref{newreheat}) should also be 
compared with $T_R \approx m_\phi/g$ which would apply in the case 
large thermal mass could forbid inflaton decay completely.

Our result has important implications to the abundances of supermassive
dark matter particles and gravitinos as well as baryon asymmetry.
First, in case a large thermal mass forbids inflaton decay completely,
the temperature after preheating cannot be higher than $\sim m_\phi/g$.
Then the abundance of supermassive particles with mass $m_X$ is
exponentially suppressed as $\propto (m_X/m_\phi)^2e^{-2m_X/m_\phi}$
for $m_X \gg m_\phi$ and $g=1$ \cite{Kolb:2003ke}.  Our result shows
that such a suppression is absent and an appreciable amount of
supermassive particles could be created after inflation.
Second, the gravitino abundance is
 suppressed by a factor of $3g^2/(24\pi^2)$ in the present model 
compared
with the conventional reheating scenario, because
its abundance is proportional to the reheat temperature.  Hence we can 
relax constraint imposed by the gravitino decay to this extent.
 
In summary, we have calculated the dissipation rate of an oscillating
scalar field in a thermal bath
such as the inflaton in the late reheating stage, and shown that it
is nonvanishing even if the would-be decay products have a thermal
mass larger than the mass of the oscillating field.  This yields
several important implications that has not been taken into account
so far, as discussed above.

\acknowledgments{ 
This work was partially supported by the JSPS
  Grant-in-Aid for Scientific Research No.\ 16340076.}
%\newpage

\end{document}